\documentstyle[11pt]{article}

\title{Spontaneous chiral-symmetry breaking of lattice QCD with massless dynamical quarks
\thanks{Supported
by the Key Project of National Science Foundation (10235040), Key
Project of National Ministry of Eduction (105135), Project of the
Chinese Academy of Sciences (KJCX2-SW-N10) and Guangdong Ministry
of Education.}}

\author{Xiang-Qian Luo\\
{\small\sl Department of Physics, Zhongshan (Sun Yat-Sen)
University,
Guangzhou 510275, China\thanks{Email: stslxq@zsu.edu.cn}}\\
}

\date{\today}

\begin{document}
\maketitle

\begin{abstract}
One of the most challenging issues in QCD is
the investigation of spontaneous chiral-symmetry breaking,
which is characterized by the non-vanishing chiral condensate when the bare fermion mass is zero.
In standard methods, one has to perform expensive lattice simulations at multiple bare quark masses, and
employ some modeled function to extrapolate the data to the chiral limit.
This paper applies the probability distribution function method to computing the chiral condensate in
lattice QCD with massless dynamical quarks, without any ambiguous mass extrapolation.
The results for staggered quarks indicates that the method
might be a more efficient alternative for investigating the spontaneous
chiral-symmetry breaking in lattice QCD.
\end{abstract}



\section{INTRODUCTION}
\label{sec1}

Chiral-symmetry breaking
is an important feature of quantum field theory with fermions.
In QCD with massless quarks at low temperature or density, the system is confined and
chiral symmetry is spontaneously broken.
At sufficiently high temperature or density, there is a chiral phase
transition from the confinement phase to the quark-gluon plasma phase
or color-superconducting phase, where chiral symmetry is restored.
There have been many phenomenological or QCD-inspired model studies
of spontaneous chiral-symmetry breaking and chiral phase
transitions, but lattice gauge theory (LGT)\cite{Wilson:1974sk}
should give the most reliable information from the first principles.
These issues\cite{Muroya:2003qs,Katz:2003up,Lombardo:2004uy,Luo:2004mc,Gregory:1999pm,Luo:2000xi,Fang:2002rk,Luo:2004se,Chen:2004tb}
have been extensively investigated on the lattice for a long time,
but there are still many unsolved problems.

The main question we would address here is how to get quantitative information
on spontaneous chiral-symmetry breaking from
lattice simulations with dynamical quarks.
When the bare fermion mass $m$ is zero, the action  is
invariant under the  global chiral transformation of the fermion field
$\psi \to \exp(i \mbox{\boldmath $\alpha \cdot \tau$}\gamma_5) \psi$,
with $\mbox{\boldmath $\tau$}$ the generator of the chiral symmetry group,
but the bilinear operator ${\bar \psi}\psi$ changes.
A non-vanishing vacuum expectation value $\langle {\bar \psi}\psi \rangle\vert_{m=0}$
is the signal for spontaneous chiral-symmetry breaking,
which is also associated with dynamical generation of fermion mass.

Suppose the quarks are degenerate. The chiral condensate per  flavor is
\begin{eqnarray}
\langle {\bar \psi} \psi \rangle
= \langle {\rm Tr} \Delta^{-1} \rangle /(N_cV),
\label{thermo}
\end{eqnarray}
where $N_c=3$ is the number of colors, $V$ is the number of
lattice sites and $\Delta$ is the fermonic matrix.
The trace is taken in the color, spin and position space.
The expectation value ``$< ...>$" is computed with the integration
measure associated with the partition function
\begin{eqnarray}
Z  &= & \int \left[ d{\bar \psi} \right] \left[ d \psi \right] \left[ dU \right]
~  \exp \left(-S \right)
=
\int \left[ dU \right] ~ \exp \left(-S_g + \ln {\rm det}
\Delta \right) , \label{partition}
\end{eqnarray}
where $U$ is the gauge link variable,
$S_g$ is the gluonic action, and $\Delta=m+i\Lambda$ is the fermionic matrix.

A popular way to calculate the chiral condensate is to use the matrix inversion technique.
Another popular way is to use
\begin{eqnarray}
\langle {\bar \psi} \psi \rangle
=
{1 \over N_cV} \bigg \langle \sum_{j=1}^{NcV/2}
{2m \over \lambda_j^2 +m^2} \bigg \rangle ,
\label{ps1}
\end{eqnarray}
with $\lambda_j$ the $j$-th positive eigenvalue of $\Lambda$.
The disadvantage is that it requires the calculation of all eigenvalues of the Dirac operator.
When the lattice volume is
large, the computational task is huge and not so
feasible. It is also very expensive to generate configurations with dynamical quarks at multiple fermion masses.
To get the chiral condensate in the chiral limit $\lim_{m \to 0} \lim_{V \to \infty}
\langle {\bar \psi} \psi \rangle$,
it requires a $m$ extrapolation, since $<{\bar \psi}\psi>\vert_{m=0}=0$ on a finite lattice.
Such an extrapolation of $\langle {\bar \psi} \psi \rangle$
to the massless limit uses some modeled fitting functions
(e.g., linear function, plus quadratic or logarithmic corrections).
Unfortunately, such a process might not be well justified, and sometimes it gives
arbitrary results, in particular when the system is at a phase transition point.

In Ref.\cite{Azcoiti:1995dq}, an alternative, named the probability distribution
function (p.d.f.) method of the chiral condensate was proposed to investigate
spontaneous chiral-symmetry breaking in LGT with fermions.
The p.d.f. method has been tested in the Schwinger model\cite{Azcoiti:1995dq},
and applied to the investigation of the spontaneous P and CT symmetry breaking\cite{Azcoiti:1999rq}, theta-vacuum like systems\cite{Azcoiti:2002vk}
and the phase transition of SU(2) LGT at finite density\cite{Aloisio:2000rb},
as well as SU(3) LGT with quenched staggered quarks\cite{Luo:2004wx}.

In this article, we will compute the chiral condensate of
lattice QCD with  dynamical staggered quarks, using the p.d.f. method.
In Sec. \ref{sec2}, we review the basic ideas.
In Sec. \ref{sec3}, we  present the numerical results.
Some discussions are made in Sec. \ref{sec4}.

\section{FORMULATION}
\label{sec2}

Let us characterize each vacuum state by $\alpha$ and
the chiral condensate  by $\langle {\bar \psi} \psi \rangle_{\alpha} $.
The p.d.f. of the chiral condensate in the Gibbs state is defined by \cite{Azcoiti:1995dq,Luo:2004wx}
\begin{eqnarray}
P(c)=\sum_{\alpha} w_{\alpha} \delta \left(c- \langle {\bar \psi} \psi \rangle_{\alpha} \right),
\label{definition}
\end{eqnarray}
with $w_{\alpha}$  the weight to get the vacuum state $\alpha$.
$P(c)$ tells us the probability to get the value $c$
for the chiral condensate from a randomly chosen vacuum state.
If the ground state is invariant under the global chiral transformation, then $P(c)=\delta(c)$.
If the ground state breaks the chiral symmetry,
$P(c)$ will be a more complex function, depending on the group of global chiral symmetry. Therefore,
from the shape of the function $P(c)$ computed
in the configurations generated by a chiral {\it symmetric} action with exact $m=0$,
one can qualitatively judge whether chiral symmetry is spontaneously
broken.

In quantum field theory with fermions,
chiral-symmetry breaking is dominated
by the properties of the fermion fields
under global chiral transformation.
From Eq. (\ref{definition}),
one can  define \cite{Luo:2004wx}
the p.d.f. of the chiral condensate
for a single gauge configuration $U$:
\begin{eqnarray}
P_U(c) ={ \int \left[ d{\bar \psi} \right] \left[ d \psi \right]
\exp \left(-S_f \right)
 \delta \left[c- {\sum_{x} {\bar \psi}(x) \psi(x)\over N_cV} \right]  \over
\int \left[ d{\bar \psi} \right] \left[ d \psi \right]
~  \exp \left(-S_f \right)} ,
\label{D1}
\end{eqnarray}
where in ``$\sum_{x} {\bar \psi}(x) \psi(x)$",
summation over color,  spin and position indices is implied.

For gauge theory with fermions, if the action is invariant under $U(1)$
chiral transformation, we show \cite{Luo:2004wx} from p.d.f. that one can use
\begin{eqnarray}
C(j)
=
\langle c_0 (U)  \rangle =
{z(j) \over N_cV}
\bigg \langle {1 \over \lambda_j} \bigg \rangle
\label{ps4}
\end{eqnarray}
to compute the chiral condensate in the exact chiral limit.
$c_0(U)$ is the amplitude
of the chiral condensate for configuration $U$ in the $m=0$ case.
$z(j)$ is the $j-$th zero of $J_0$, with $J_0$
the zeroth order Bessel function of the first kind.
The average is taken over gauge configurations with dynamical fermions at $m=0$.
In the chiral-symmetry breaking phase,
a plateau for $C(j) = const.$ will develop, from which the chiral condensate
in the chiral limit can be extracted.

The relation between eigenmodes
and chiral-symmetry breaking is clear:
if chiral symmetry
is spontaneously broken, i.e.,
$c_0(U) \not\equiv 0$, according to Eq. (\ref{ps4}),  $\lambda_j$
should scale as $z(j)/V$.
In the infinite volume limit $V\to \infty$,  the eigenvalues relevant for
chiral-symmetry breaking are those going to zero as
$1/V$, which is consistent with Banks-Casher.

The advantage of Eq. (\ref{ps4}) is that to extract the value of $C(j)$ from a plateau,
only a few smallest eigenvalues for configurations with dynamical quarks at $m=0$ are needed
for this calculation. Of course,
finite size analysis
$\lim_{V \to \infty} C(j)$
remains to be done, as in
all approaches.

\section{RESULTS FOR QCD WITH DYNAMICAL QUARKS}
\label{sec3}

In Ref. \cite{Luo:2004wx}, we have displayed how the method works in quenched lattice QCD.
Here we would like to present the first data for SU(3) LGT
with dynamical staggered (Kogut-Susskind) quarks.
The action is $S=S_g+S_f$:
\begin{eqnarray}
S_g &=& - {\beta \over N_c} \sum_{p} {\rm Re} ~ {\rm Tr} (U_p) ,
\nonumber \\
S_f &=& \sum_{x,y} {\bar \psi}(x) \Delta_{x,y} \psi (y) ,
\nonumber \\
U_p &=& U_{\mu}(x) U_{\nu}(x+\mu) U_{\mu}^{\dagger} (x+\nu) U_{\nu}^{\dagger} (x) ,
\nonumber \\
\Delta_{x,y}  &=& m \delta_{x,y}
+
\sum_{\mu=1}^4 {1 \over 2} \eta_{\mu}(x)  \bigg[ U_{\mu}(x) \delta_{x,y-{\hat \mu}}
- U_{\mu}^{\dagger} (x-{\hat \mu}) \delta_{x,y+{\hat \mu}} \bigg] ,
\nonumber \\
\eta_{\mu}(x) &=& (-1)^{x_1+x_2+...+x_{\mu-1}} ,
\end{eqnarray}
where $\beta=2 N_c/g^2$, with $N_c=3$.
In the chiral limit $m=0$,
there exists a U(1) subgroup of the continuous chiral symmetry in the fermionic action and
Eq. (\ref{ps4}) applies.
All simulations are done on the $V=10^4$ lattice using the method described in
\cite{Luo:2001id}.

Figure \ref{fig1} shows the data for $\langle {\bar \psi} \psi \rangle$ for flavor number $N_f=1$ at $\beta=5.1069$ obtained from Eq. (\ref{ps1}) at 14 nonzero $m$.
For conventional simulation methods,
it is very expensive to do computations at so many quark masses.
A linear function is also used to extrapolate the data of $\langle {\bar \psi} \psi \rangle$ at $m \neq 0$ to the chiral limit.
 At this coupling, the linear fit works well for wide range of $m \in [0.004,0.1]$, so that the result can be used for
comparison.

Figure \ref{fig2} shows the result of $C(j)$ for the same
$\beta$ and $N_f$ using Eq. (\ref{ps4}) and from only one simulation\cite{Luo:2001id} at $m=0$.
There is a nice
plateau for $j \in [5,40]$. The result is consistent with that from Eq. (\ref{ps1}) extrapolated to the chiral limit and Eq. (\ref{ps4}).

Figures \ref{fig3}-\ref{fig8} show respectively the results for $N_f=2$ at $\beta=5.0238$,
$N_f=3$ at $\beta=4.9411$, and $N_f=4$ at $\beta=4.8590$. In the chiral limit $m=0$,
the data for different flavors have the same plaquette energy $E_p=1.2921$, but different $\beta$,
due to the effects of dynamical quarks.
Our results indicate that the p.d.f. method gives estimate for the chiral condensate,
consistent with the conventional one.

\section{DISCUSSIONS}
\label{sec4}

To summarize,
we have extended the p.d.f. method to lattice QCD with dynamical staggered quarks
for computing the chiral condensate in the exact chiral limit.
The first results are very encouraging.
This might be an alternative efficient method for obtaining quantitative information on  the spontaneous
chiral-symmetry breaking in QCD.
There are several advantages:
only calculations of a small set of eigenvalues of the massless Dirac operator
are necessary; only
one simulation at $m=0$ has to been done at one $\beta$; there is no need for $m$ or $\lambda$ extrapolation.

In some cases, i.e., when the system is at the criticality, fitting the data
obtained from obtained from Eq. (\ref{ps1}) with some modeled function might give
arbitary results, while using Eq. (\ref{ps4}) might hopefully give more reliable estimate.

\vskip 1cm

\noindent
{\bf Acknowledgments}

I am grateful to V. Azcoiti and V. Laliena for useful discussions.



\begin{figure}[htb]
\begin{center}
\input pm_nf1_b5p1069.tex
\end{center}
\caption{$\langle \bar{\psi} \psi \rangle /4$ as a function of $m$
for $N_f=1$ and $\beta=5.1069$
using Eq. (\ref{ps1}).}
\label{fig1}
\end{figure}

\begin{figure}[htb]
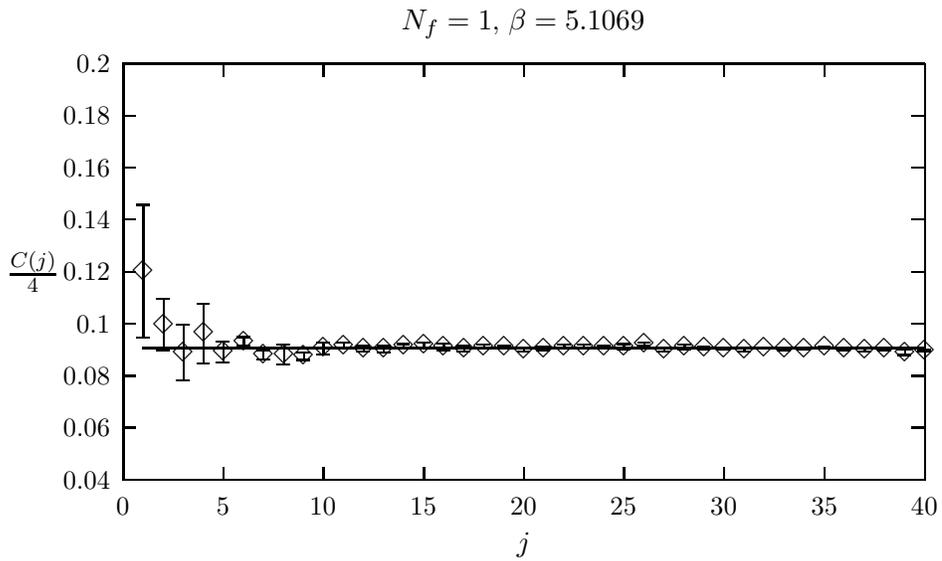

\begin{center}
\input zero_nf1_e1p2921.tex
\end{center}
\caption{$C(j)/4$ as a function of $j$
for $N_f=1$ and $\beta=5.1069$ using Eq. (\ref{ps4}).}
\label{fig2}
\end{figure}

\begin{figure}[htb]
\begin{center}
\input pm_nf2_b5p0238.tex
\end{center}
\caption{$\langle \bar{\psi} \psi \rangle /4$ as a function of $m$
for $N_f=2$ and $\beta=5.0238$
using Eq. (\ref{ps1}).}
\label{fig3}
\end{figure}

\begin{figure}[htb]
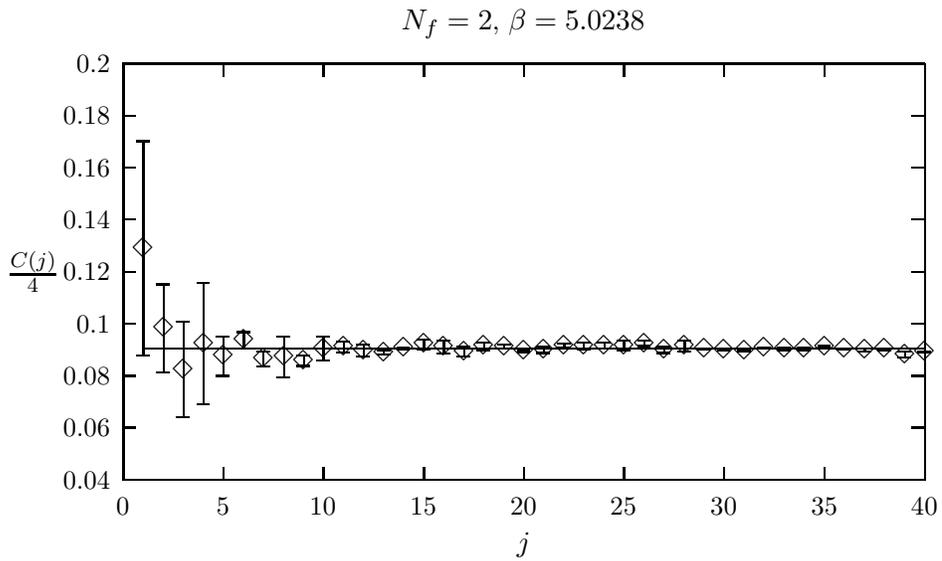

\begin{center}
\input zero_nf2_e1p2921.tex
\end{center}
\caption{$C(j)/4$ as a function of $j$
for $N_f=2$ and $\beta=5.0238$ using Eq. (\ref{ps4}).}
\label{fig4}
\end{figure}

\begin{figure}[htb]
\begin{center}
\input pm_nf3_b4p9411.tex
\end{center}
\caption{$\langle \bar{\psi} \psi \rangle /4$ as a function of $m$
for $N_f=3$ and $\beta=4.9411$
using Eq. (\ref{ps1}).}
\label{fig5}
\end{figure}

\begin{figure}[htb]
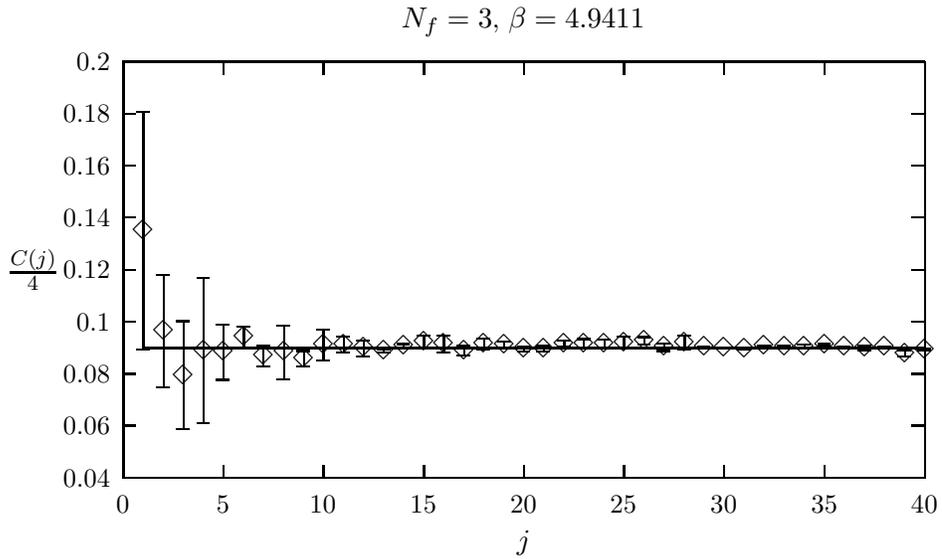

\begin{center}
\input zero_nf3_e1p2921.tex
\end{center}
\caption{$C(j)/4$ as a function of $j$
for $N_f=3$ and $\beta=4.9411$ using Eq. (\ref{ps4}).}
\label{fig6}
\end{figure}

\begin{figure}[htb]
\begin{center}
\input pm_nf4_b4p8590.tex
\end{center}
\caption{$\langle \bar{\psi} \psi \rangle /4$ as a function of $m$
for $N_f=4$ and $\beta=4.8590$
using Eq. (\ref{ps1}).}
\label{fig7}
\end{figure}

\begin{figure}[htb]
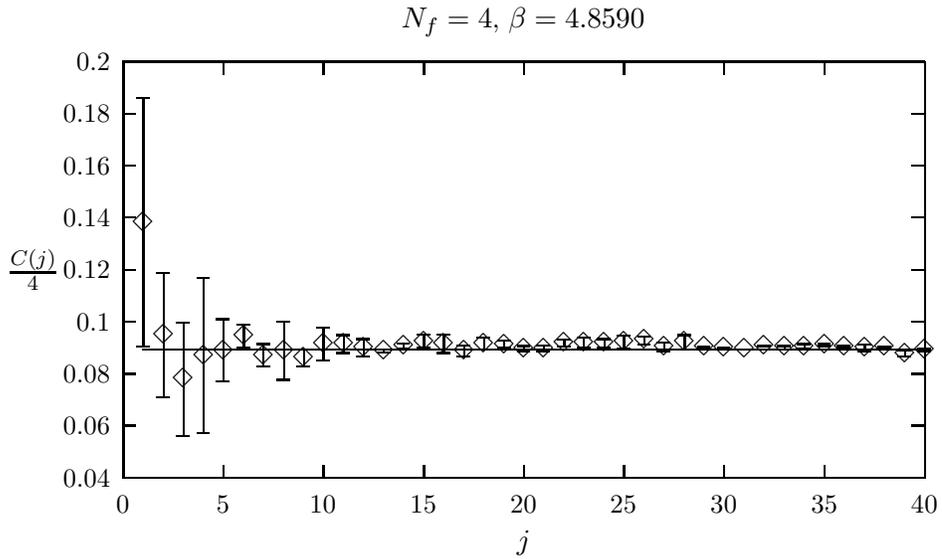

\begin{center}
\input zero_nf4_e1p2921.tex
\end{center}
\caption{$C(j)/4$ as a function of $j$
for $N_f=4$ and $\beta=4.8590$ using Eq. (\ref{ps4}).}
\label{fig8}
\end{figure}

\end{document}